\documentclass[conference]{IEEEtran}
\IEEEoverridecommandlockouts

\usepackage{color}
\usepackage{times}
\usepackage{cite}
\usepackage{flushend}
\usepackage{supertech-acm}
\usepackage{xspace} 
\usepackage{wrapfig}

\usepackage[multiple]{footmisc}
\usepackage{pgfplots, pgfplotstable}
\usetikzlibrary{patterns}
\usepackage{geometry}
\usepackage{hyperref}
\hypersetup{
    colorlinks=true,
    linkcolor=black,
    citecolor=black,
}

\newcommand{\work}{T_1}
\newcommand{\spa}{T_\infty}
\newcommand{\serial}{T_S}
\newcommand{\NUMAWS}{NUMA-WS\xspace} 
\newcommand{\spawn}{\texttt{cilk\_spawn}\xspace}
\newcommand{\sync}{\texttt{cilk\_sync}\xspace}
\newcommand{\cilkfor}{\texttt{cilk\_for}\xspace}

\newcommand{\anote}[1]{\cnote{red}{Angelina: #1}}

\newcounter{groupcount}
\begin{document}

 \begin{titlepage}
© 2018 IEEE. Personal use of this material is permitted. Permission from IEEE must be obtained for all other uses, in any current or future media, including reprinting/republishing this material for advertising or promotional purposes, creating new collective works, for resale or redistribution to servers or lists, or reuse of any copyrighted component of this work in other works.\
\\
\\
You can find the published work on IEEE Xplore by following this \href{https://ieeexplore.ieee.org/document/8573486}{link}.
\end{titlepage}

\title{A NUMA-Aware Provably-Efficient Task-Parallel Platform Based
on the Work-First Principle
\thanks{This research was supported in part by National Science Foundation under 
grant number CCF-1527692 and~CCF-1733873.}}

\author{
\IEEEauthorblockN{
Justin Deters \hspace{3em}
Jiaye Wu \hspace{3em}
Yifan Xu \hspace{3em}
I-Ting Angelina Lee}
\vspace{1mm}
\IEEEauthorblockA{
\textit{Washington University in St. Louis}\\
\{j.deters, jiaye.wu, xuyifan, angelee\}@wustl.edu}
}

\maketitle
\notesfalse

\begin{abstract}

Task parallelism is designed to simplify the task of parallel programming.
When executing a task parallel program on modern NUMA architectures, it can
fail to scale due to the phenomenon called \defn{work inflation}, where the
overall processing time that multiple cores spend on doing useful work is 
higher compared to the time required to do the same amount of work on one 
core, due to effects experienced only during parallel executions such
as additional cache misses, remote memory accesses, and memory bandwidth
issues.

One can mitigate work inflation by co-locating the computation with
its data, but this is nontrivial to do with task parallel programs.  First, by
design, the scheduling for task parallel programs is automated, giving the
user little control over where the computation is performed.  Second, the
platforms tend to employ work stealing, which provides strong theoretical
guarantees, but its randomized protocol for load balancing does not discern
between work items that are far away versus ones that are closer.

In this work, we propose \NUMAWS, a NUMA-aware task parallel platform
engineered based on the work-first principle.  By abiding by the work-first
principle, we are able to obtain a platform that is work efficient, provides
the same theoretical guarantees as a classic work stealing scheduler, and
mitigates work inflation.  We have extended Cilk Plus runtime system to 
implemented \NUMAWS.  Empirical results indicate that the \NUMAWS is work 
efficient and can provide better scalability by mitigating work inflation.  

\end{abstract}

\begin{IEEEkeywords}
work stealing, work-first principle, NUMA, locality, work inflation
\end{IEEEkeywords}

\secput{intro}{Introduction}

Modern concurrency platforms are designed to simplify the task of writing
parallel programs for shared-memory parallel systems.  These platforms
typically employ \defn{task parallelism} (sometimes referred to as 
\defn{dynamic multithreading}), in which the programmer expresses the \emph{logical}
parallelism of the computation using high-level language or library constructs
and lets the underlying scheduler determine how to best handle synchronizations
and load balancing.  Task parallelism provides a programming model that is
\defn{processor oblivious}, because the language constructs expose the logical
parallelism within the application without specifying the number of cores on
which the application will run.  Examples of such platforms include
OpenMP~\cite{OpenMP13}, Intel's Threading Building Blocks
(TBB)~\cite{Reinders07,IntelTBBManual}, various Cilk
dialects~\cite{BlumofeJoKu96,FrigoLeRa98,DanaherLeLe06,LeeBoHu10,Leiserson10,
IntelCilkPlus13}, various Habanero dialects~\cite{BarikBuCa09,CaveZhSh11},
Java Fork/Join Framework~\cite{Lea00},	and IBM's X10~\cite{CharlesGrSa05}.

Most concurrency platforms, including ones mentioned above, schedule
task parallel computations using \defn{work stealing}~\cite{BlumofeLe94,
BlumofeLe99, AroraBlPl98, AroraBlPl01}, a randomized distributed protocol for
load balancing.  Work stealing, in its classic form, provides strong theoretical
guarantees.  In particular, it provides asymptotically optimal execution
time~\cite{BlumofeLe94,BlumofeLe99,AroraBlPl98,AroraBlPl01} and allows for
good cache locality with respect to sequential execution when using private
caches~\cite{AcarBlBl00}.  In practice, work stealing has also been
demonstrated to incur little scheduling overhead and can be implemented
efficiently~\cite{FrigoLeRa98}.  

\begin{wrapfigure}{r}{0.48\columnwidth}
\vspace{-3mm}
\includegraphics[width=4.2cm]{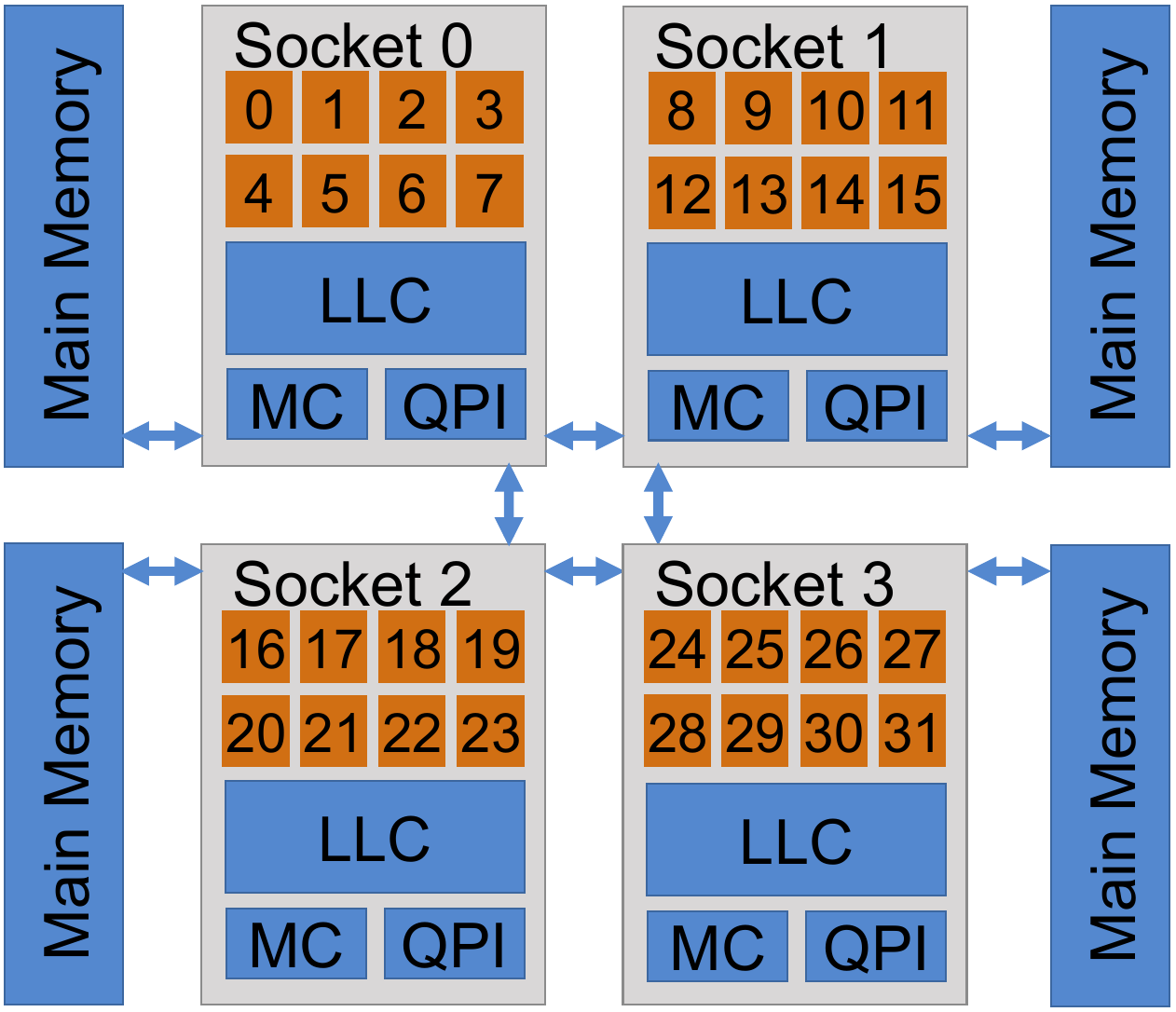}
\vspace{-2mm}
\caption{An example of a 32-core four-socket system, where each socket has its
own last-level L3 cache and memory banks.}
\vspace{-2mm}
\label{fig:numa}
\end{wrapfigure}

Shared memory on modern parallel systems is often realized with
\defn{Non-Uniform Memory Access (NUMA)}, where the memory latency can vary
drastically, depending on where the memory access is serviced.  \figref{numa}
shows an example of a NUMA system and its memory subsystem.  This system
consists of four sockets, with eight cores per socket, and each socket has its
own last-level cache (LLC), memory controller, and memory banks (DRAM).  Each
LLC is shared among cores on the same socket, and the main memory consists of
all the DRAMs across sockets, where each DRAM is responsible for a subset of
the physical address range.  On such a system, when a piece of data is allocated,
it can either reside on physical memory managed by the \defn{local} DRAM
(i.e., on the same socket) or by the \defn{remote} DRAM (i.e., on a different
socket).  When accessed, the data is brought into the local LLC and its
coherence is maintained by the cache coherence protocol among LLCs.  Thus, the
memory access latency can be tens of cycles (serviced from the local LLC),
over a hundred cycles (serviced from a local DRAM or a remote LLC), or a few
hundreds of cycles (serviced from a remote DRAM).

A task parallel program can fail to scale on such a NUMA system, as a result
of a phenomenon called \defn{work inflation}, where the overall processing
time that multiple cores spend on doing useful work is much higher compared to
the time required to do the same amount of work on one core, due to effects
experienced only during parallel executions.  Multiple factors can contribute
to work inflation, including work migration, parallel computations sharing a
LLC destructively, or accessing data allocated on remote sockets. 

One can mitigate work inflation by co-locating computations that share the
same data or co-locate the computation and its data on the same socket,
thereby reducing remote memory accesses.  These strategies are not
straightforward to implement in task parallel code scheduled using work
stealing, however.  First, by design the scheduling of task parallel programs
is automated, which gives the programmer little control over where the
computation is executed.  Second, the randomized protocol in work stealing
does not discern between work items that are far away versus ones that are
closer.

Ideally, we would like a task parallel platform that satisfies the following
criteria:
\begin{closeitemize}
\item provide the same strong theoretical guarantees that a classic work 
    stealing scheduler enjoys;
\item be \defn{work efficient}, namely, the platform does not unnecessarily 
    incur scheduling overhead that causes the single-core execution time 
    to increase;
\item support a similar processor-oblivious model of computation: assuming 
    sufficient parallelism, the same program for a given input
    should scale as the number of cores used increases; and
\item mitigate work inflation.
\end{closeitemize}
Even though many mechanisms and
scheduling policies have been proposed to mitigate work inflation in task
parallel programs~\cite{GuoZhCa10, ChenHuGu11, LifflanderKrKa14,
OlivierPoWh12, ChenGuHu12, OlivierDeSuSc12, MuddukrishnaJoVl13,
SimhadriBlFi14, DrebesHeDr14, DrebesPoHe16, MajoGr17}, none of the proposed
solutions satisfy all criteria simultaneously.  In particular, many of them
are not work efficient nor do they provide a provably efficient scheduling 
bound (see \secref{related}).

In this paper, we propose \NUMAWS, a task parallel platform that satisfies
these criteria simultaneously.  \NUMAWS employs a variant of work stealing
scheduler that extends the classic algorithm with mechanisms to mitigate NUMA
effects.  \NUMAWS achieves the same theoretical bounds on execution time and
additional cache misses for private caches as the classic work stealing
algorithm, albeit with a slightly larger constant hidden in the big-$O$ term.

\NUMAWS provides the same execution time bound as classic work
stealing~\cite{BlumofeLe94,BlumofeLe99,AroraBlPl98,AroraBlPl01}, which can be
quantified using two important metrics: the \defn{work}, as defined by the
execution time running the computation on one core, and the \defn{span}, the
longest sequential dependences in the computation, or its theoretical running
time executing on infinitely-many cores.  Given a computation with $\work$
\defn{work} and $\spa$ \defn{span}, \NUMAWS executes the computation on $P$
cores in expected time $\work / P + O(\spa)$.\footnote{Even without accounting
for scheduling overhead, this is the best bound possible when the dependences
of the parallel computation are not unknown until execution
time~\cite{Brent74,Graham69}.}  The additional cache misses due to parallel
execution are directly correlated with the number of times computation
``migrates,'' or when the order of computation during parallel execution
diverges from that of a single-core execution.  In \NUMAWS, the number of
times such divergence can occur is upper bounded by $O(P\spa)$, same as with
classic work stealing~\cite{AcarBlBl00}. 

To measure work efficiency, an important metric is $\serial$, the execution
time of \defn{serial elision}, obtained by removing the parallel control
constructs or replacing it with its sequential counter part.  Serial elision
should perform the exact same algorithm that the parallel program implements
but without the parallel overhead.  Thus, one can quantify work efficiency by
comparing $\work$ against $\serial$ to measure parallel overhead.  Assuming a
work-efficient platform, one can obtain parallel code whose ratio between
$\work$ and $\serial$ is close to one.

We have implemented a prototype system by extending Intel Cilk Plus,
which implements the classic work stealing algorithm, and empirically
evaluated it.  The empirical results indicate that \NUMAWS is work efficient,
scalable across different number of cores, and can mitigate NUMA effects.
Specifically, we show that, \NUMAWS incurs negligible parallel overhead (i.e.,
$\work/\serial$), comparable to that in Cilk Plus.  Moreover, we compare the
parallel execution times across benchmarks when running on Intel Cilk Plus
versus running on \NUMAWS on a four-socket 32-core system.  \NUMAWS was able
to decrease work inflation compared to Cilk Plus without adversely impacting
scheduling time.  Across benchmarks, \NUMAWS obtains much better speedup than
that in Cilk Plus.

Critically, to achieve the theoretical bounds and practical efficiency, the
design and engineering of \NUMAWS abides by a principle called the
\defn{work-first principle} proposed by~\cite{FrigoLeRa98}, which states
that one should minimize the overhead borne by the work term ($\work$) and
move the overhead onto the span term ($\spa$).  Intuitively, a scheduler must
incur some scheduling overhead due to the extra bookkeeping necessary to
enable correct parallel execution or to mitigate NUMA effects.  Within the
context of a work-stealing scheduler, worker threads (surrogates of processing
cores) load balance by ``stealing'' work when necessary.  The work-first
principle states that it's best to incur scheduling overhead on the control
path that can be amortized against successful steals.  To put it differently,
whenever a choice can be made to incur overhead on a thief stealing versus on
a worker busy working, it's always preferred to incur the overhead on the
thief stealing. 

\subsection*{\bf Contributions} 

To summarize, this paper makes the following contributions: 
\begin{closeitemize}
\item We present \NUMAWS, a NUMA-aware task parallel platform that implements
a work stealing algorithm with mechanisms to mitigate work inflation
(\secref{numaws}).

\item We show that our extended work stealing algorithm retains the
same theoretical guarantees on execution time and cache bounds as the 
classic algorithm (\secref{analysis}).

\item We implemented and empirically evaluated \NUMAWS.  The empirical results
show that \NUMAWS is work efficient, scalable across cores, and mitigates work
inflation (\secref{eval}).  
\end{closeitemize}

\punt{ The rest of the paper is organized as follows.  \secref{motivation}
motivates the problem by examining Cilk Plus, the system that \NUMAWS is based
on.  We show that Cilk Plus is work efficient but can fail to scale due to
work inflation.  \secrefs{numaws}{eval} present the system, its analysis, and
evaluation.  Finally, \secref{related} discusses  prior work on NUMA-aware
task parallel platforms and why they failed to satisfy our desired goals.  }
\secput{motivation}{Preliminaries: Work Stealing in Cilk Plus} 

Our prototype implementation of \NUMAWS extends Intel Cilk
Plus~\cite{IntelCilkPlus13}, which implements the classic work stealing
algorithm.  The engineering of Cilk Plus also follows the work-first
principle.  In this section, we review the implementation of Cilk Plus and
examine its work efficiency to demonstrate the benefit of the work-first
principle.  Next, we examine the work inflation of multiple benchmarks running
on Cilk Plus to motivate the need for a NUMA-aware work-efficient runtime.


\paragraph{The language model.} Cilk Plus extends C/C++ with two parallel
primitives: \spawn and \sync.\footnote{A third keyword \cilkfor exists, which 
specifies that the iterations for a given loop can be executed in parallel; 
it is syntactic sugar that compiles down to binary spawning of iterations 
using \spawn and \sync.  Other concurrency platforms contain
similar constructs with similar semantics, though the syntax may differ
slightly.}  When a function $F$ \defn{spawns} another function $G$ (invoking
$G$ with the keyword \spawn), the \defn{continuation} of $F$, i.e., statements
after the \spawn call, may execute in parallel with $G$.  The keyword \sync
specifies that control cannot pass beyond the \sync statement until all
previously spawned children within the enclosing function have
returned.

These keywords denote the \emph{logical} parallelism of the computation.  When
$F$ spawns $G$, $G$ may or may not execute in parallel with the continuation
of $F$, depending on the hardware resource available during execution.  

\paragraph{Work stealing and the work-first principle.} In work stealing, each
\defn{worker} (a surrogate of a processing core) maintains a \defn{deque} (a
double ended queue) of work items.  Each worker operates on its own deque
locally most of the time and communicates with one another only when it runs
out of work to do, i.e., its deque becomes empty.  When that happens, a worker
turns into a \defn{thief} and \emph{randomly} chooses another worker, the
\defn{victim}, to steal work from.  A worker, while busy working, always 
operates at the tail of its own deque like a stack (i.e., first in last out).
A thief, when stealing, always steals from the head of a victim's deque 
(i.e., taking the oldest item).

The \defn{work-first} principle~\cite{FrigoLeRa98} states that one should
minimize the overhead borne by the work term ($\work$) and move the overhead
onto the span term ($\spa$), which corresponds to the steal path.  A
work-stealing runtime abiding by the work-first principle tends to be work
efficient, as demonstrated by the implementation of Cilk-5.  Subsequent
variants of Cilk~\cite{DanaherLeLe06, LeeBoHu10, Leiserson10} including Cilk
Plus follow similar design.

The intuition behind the work-first principle can be understood as follows.
The parallelism of an application is defined as $\work/\spa$, or how much work
there is along each step of the span.  Assuming the application contains ample
parallelism, i.e., $\work/\spa \gg P$, the execution time
is dominated by the $\work / P$ term, and thus it's better to incur overhead
on the $\spa$ term.  Moreover, in practice, when the application contains
ample parallelism, steals occur infrequently.

\newcommand{\negspace}{-3mm}
\begin{figure}
\footnotesize
\begin{tabular}{|l|}
\hline
\begin{minipage}[t]{0.95\linewidth}
\vspace{\negspace}
\begin{codebox}
\Procname{$\id{F}$ spawns $\id{G}$:}
\li    $\proc{pushDequeAtTail}$($\id{F}$); \label{li:spawn}
            \Comment{$\id{F}$'s continuation becomes stealable}
\li    continue to execute $\id{G}$ \label{li:exe-child}
\end{codebox}
\end{minipage}%
\\ \hline
\begin{minipage}[t]{0.95\linewidth}
\vspace{\negspace}
\begin{codebox*}
\Procname{$\id{G}$ returns to its spawning parent $\id{F}$:}
\li    $\id{success}$ = $\proc{popDequeAtTail}$(); \label{li:spawn-ret}
\li    \If $\id{success}$ \label{li:pop-succ}
            \Comment{$\id{F}$ is not stolen and must be at the tail of the deque }
\li    \Then continue to execute $\id{F}$ \label{li:cont}
\li    \Else \label{li:return-else}
            \Comment{parent stolen; the deque is empty}
\li       $\id{next\_action}$ = $\const{CHECK\_PARENT}$ \label{li:check-parent}
\li       return to scheduling loop \label{li:return-loop}
      \End
\end{codebox*}
\end{minipage}%
\\ \hline
\begin{minipage}[t]{0.95\linewidth}
\vspace{\negspace}
\begin{codebox*}
\Procname{$\id{F}$ executes \sync:}
\li \If $\attrib{F}{stolen}$ = $\const{true}$ 
\li    \Then \Comment{$\id{F}$ must be a full frame and the deque is empty}
\li    $\id{success}$ = $\proc{checkSync}$(); \label{li:check-sync}
       \Comment{must do a nontrivial sync}
\li    \If $\id{success}$ \Then
\li       continue to execute $\id{F}$
\li    \Else \label{li:sync-else}
            \Comment{$\id{F}$ must be the only thing in the deque}
\li       suspend $\id{F}$ \label{li:suspend}
\li       $\id{next\_action}$ = $\const{STEAL}$
\li       return to scheduling loop \label{li:sync-loop}
       \End 
\li \Else continue to execute $\id{F}$ \Comment{nothing else needs to be done}
                    \label{li:noop} 
    \End
\end{codebox*}
\end{minipage}%
\\ \hline
\begin{minipage}[t]{0.95\linewidth}
\vspace{\negspace}
\begin{codebox*}
\Procname{scheduling loop:
\Comment{$\id{frame}$ is a either $\const{null}$ or the first root full frame}}
\li   \While $\id{computation-done}$ = $\const{false}$ \Do
\li      \If $\id{next\_action}$ = $\const{CHECK\_PARENT}$ \Then \label{li:loop-check}
\li         $\id{frame}$ = $\proc{checkParent}$(); \label{li:loop-check-parent}
\li         $\id{next\_action}$ = $\const{STEAL}$ \label{li:loop-next}
                    \Comment{reset $\id{next\_action}$}
         \End
\li      \If $\id{frame}$ = $\const{null}$ \Then
\li          $\id{frame}$ = $\proc{randomSteal}$(); \label{li:steals}
\li      \Else $\proc{resume}$($\id{frame}$) \End
\end{codebox*}
\end{minipage}%
\\ \hline
\end{tabular}
\caption{Pseudocode for the Cilk Plus work-stealing scheduler: when a 
function spawns, when a spawned function returns, when a function 
executes \sync, and its scheduling loop.  Here, we use $\id{F}$ to 
represent both a function instance and its corresponding frame. 
The variable $\id{next\_action}$ specifies 
what the scheduling loop should do next.} 
\label{fig:code-cilkplus}
\vspace{-4mm}
\end{figure}

\paragraph{Work stealing in Cilk Plus.} \figref{code-cilkplus} shows the
pseudocode for the work-stealing scheduler in Cilk Plus.  Note that when no
steal occurs, the one-worker execution follows that of the serial elision.
Upon a \spawn, the worker pushes the \emph{continuation} of the spawning
parent at the tail of its deque (\liref{spawn}) and continues to execute
the spawned child (\liref{exe-child}), which can also spawn.  Once pushed, the
continuation of the parent becomes stealable.  Upon returning from a \spawn,
the worker pops the parent off the tail of its deque (if not stolen)
to resume its execution (\lirefs{spawn-ret}{cont}).  

The strategy of pushing the continuation of the parent is called
\defn{continuation-stealing}.  An alternative implementation is to push the
spawned child, called \defn{child-stealing}.\footnote{In the literature,
continuation-stealing is sometimes referred to as \defn{work-first} and
child-stealing referred to as \defn{help-first}.  The choice of which strategy
to implement is orthogonal to the work-first principle. Hence, we call them
``continuation'' versus ``child-stealing'' here to avoid confusion.} Cilk Plus
implements continuation-stealing because it can be more space efficient; more
importantly, it allows a worker's execution between successful steals to
mirror exactly that of the serial elision.  Thus, one can optimize the cache
behavior of parallel code for private caches by optimizing that of the serial
elision.

\paragraph{Runtime organization based on the work-first principle.} Two
aspects of the Cilk Plus design follow from the work-first principle: the
``THE protocol,'' proposed by~\cite{FrigoLeRa98} and the organization of the
runtime data structures, as described in~\cite{FrigoHaLe09}.  
The THE protocol is designed to minimize the overhead of a worker operating on
its deque, allowing a victim who is doing work to not synchronize with a thief unless
they are both going after the same work item in the deque.  The THE protocol
remains unchanged in \NUMAWS and thus we omit the details here and refer
interested readers to \cite{FrigoLeRa98}.  We briefly review the organization
of the runtime data structures, which is most relevant to the design of
\NUMAWS.  

The runtime data structures in Cilk Plus are organized around the work-first
principle, so as to incur as little overhead on the work path as possible, at
the expense of incurring overhead on the steal path.  With
continuation-stealing, in the absence of any steals, the behavior of a worker
mirrors exactly that of the serial elision, and the execution should incur
little scheduling overhead.  On the other hand, when a successful steal
occurs, \emph{actual} parallelism is realized, because a successful steal
enables the continuation of a spawned parent to execute concurrently (on the
thief) with its spawned child (on the victim).  In this case, the runtime must
perform additional bookkeeping in order to keep track of actual parallel
execution. 

In Cilk Plus, a \defn{Cilk function} that contains parallel keywords is
treated as an unit of scheduling: every Cilk function has an associated
\defn{shadow frame} that gets pushed onto the deque upon spawning.  It is
designed to be light weight, storing the minimum amount of information necessary
in order to enable parallel execution (i.e., which continuation to resume
next).  Whenever a frame is stolen successfully, however, the runtime promotes
the stolen frame from a shadow frame into a \defn{full frame} which contains
the necessary bookkeeping information to keep track of actual parallel
execution.  

That means, only a full frame can have spawned children executing
concurrently.  Thus, a frame's stolen field (e.g., $\attrib{F}{stolen}$) is
only ever set for a full frame that has been stolen but has not executed a
\sync.  Execution of a \sync checks for the flag, and only if the flag is set,
then a \defn{nontrivial sync} needs to be invoked that checks for outstanding
spawned children executing on other workers (\liref{check-sync}).  On the
other hand, for a shadow frame, its flag is never set and executing a \sync is
a no-op, as its corresponding function cannot have outstanding spawned
children and thus nothing needs to be done (\liref{noop}). 

If a nontrivial sync is necessary, and there are outstanding spawned
children executing on other workers, then the current worker suspends this
frame and returns to the runtime to find more work to do (i.e., steal)
(\lirefs{suspend}{sync-loop}).  The suspended frame then becomes the
responsibility of the worker who executes the last spawned child returning.
Thus, a child returning from a spawn, upon detecting that its parent has been
stolen, returns back to the scheduling loop
(\lirefs{return-else}{return-loop}) and checks if its parent is ready to resume
(\lirefs{loop-check}{loop-next}), i.e., it is the last spawned child
returning.  

Since work stealing always steals from the head of the deque, when a worker is
about to return control back to the scheduling loop
(\lireftwo{return-loop}{sync-loop}), its deque must be empty.  Thus, the
scheduling loop handles only full frames.  Upon returning to the loop, a
worker executes the loop repeatedly until the computation is done or some work
is found, either via resuming a suspended parent (\liref{loop-check-parent})
or via successful steals (\liref{steals}).

\begin{figure}[h]
\pgfplotsset{
    draw group line/.style n args={5}{
        after end axis/.append code={
            \setcounter{groupcount}{0}
            \pgfplotstableforeachcolumnelement{#1}\of\datatable\as\cell{%
                \def\temp{#2}
                \ifx\temp\cell
                    \ifnum\thegroupcount=0
                        \stepcounter{groupcount}
                        \pgfplotstablegetelem{\pgfplotstablerow}{X}\of\datatable
                        \coordinate [yshift=#4] (startgroup) at (axis cs:\pgfplotsretval,0);
                    \else
                        \pgfplotstablegetelem{\pgfplotstablerow}{X}\of\datatable
                        \coordinate [yshift=#4] (endgroup) at (axis cs:\pgfplotsretval,0);
                    \fi
                \else
                    \ifnum\thegroupcount=1
                        \setcounter{groupcount}{0}
                        \draw [
                            shorten >=-#5,
                            shorten <=-#5
                        ] (startgroup) -- node [anchor=base, yshift=0.7ex, font=\Large] {#3} (endgroup);
                    \fi
                \fi
            }
            \ifnum\thegroupcount=1
                        \setcounter{groupcount}{0}
                        \draw [
                            shorten >=-#5,
                            shorten <=-#5
                        ] (startgroup) -- node [anchor=base, yshift=0.7ex, font=\Large] {#3} (endgroup);
            \fi
        }
    }
}

\begin{tikzpicture}[scale=0.5]
\pgfplotstableread{
X	Gp	Name	Work	Scheduling	Idle
1	cilksort	P=1	1.00	0.00	0.00
2	cilksort	P=32	1.55	0.02	0.01
3	heat	P=1	0.99	0.00	0.00
4	heat	P=32	5.21	0.01	0.04
5	strassen	P=1	0.99	0.00	0.00
6	strassen	P=32	1.49	0.00	0.00
7	hull1	P=1	1.01	0.00	0.00
8	hull1	P=32	4.10	0.02	0.02		
9	hull2	P=1	1.02	0.00	0.00
10	hull2	P=32	2.32	0.00	0.01		
11	cg	P=1	1.07	0.00	0.00
12	cg	P=32	2.50	0.03	0.06		
13	matmul	P=1	1.00	0.00	0.00
14	matmul	P=32	1.09	0.00	0.00		
}\datatable

\begin{axis}[
    axis lines*=left, ymajorgrids,
    ymin=0,
    ybar stacked,
    bar width=12pt,
    width=6.5in,
    height=4in,
    legend style={font=\Large},
    draw group line={Gp}{cilksort}{cilksort}{-4ex}{10pt},
    draw group line={Gp}{heat}{heat}{-4ex}{10pt},
    draw group line={Gp}{strassen}{strassen}{-4ex}{10pt},
    draw group line={Gp}{lu}{lu}{-4ex}{10pt},
    draw group line={Gp}{hull1}{hull1}{-4ex}{10pt},
    draw group line={Gp}{hull2}{hull2}{-4ex}{10pt},
    draw group line={Gp}{cg}{cg}{-4ex}{10pt},
    draw group line={Gp}{matmul}{matmul}{-4ex}{10pt},
    xtick=data,
    xticklabels from table={\datatable}{Name},
    xticklabel style={rotate=90,xshift=-5ex,anchor=mid east,font=\Large},
    yticklabel style={font=\Large},
    ylabel style={font=\Large},
    ylabel=Total processing time normalized to $T_S$
]

\addplot table [x=X, y=Work] {\datatable}; \addlegendentry{Work,}
\addplot [fill=black] table [x=X, y=Scheduling] {\datatable}; \addlegendentry{Scheduling}
\addplot [fill=gray] table [x=X, y=Idle] {\datatable}; \addlegendentry{Idle}

\end{axis}
\end{tikzpicture}
\vspace{-4mm}
\caption{The normalized total processing times of benchmarks running on Cilk
Plus, normalized to $\serial$, the execution time of the corresponding serial
elision.  The \texttt{P=1} bars show the normalized total processing times
running on one core; the \texttt{P=32} bars show the normalized total
processing times running on $32$ cores.  For \texttt{P=32}, each data point is
also broken down into three categories: work time, scheduling time, and idle
time.  We have two data sets with different characteristics for \texttt{Hull}
and thus two sets of data points are shown for \texttt{Hull}.} 
\label{fig:cilkplus}
\vspace{-4mm}
\end{figure}
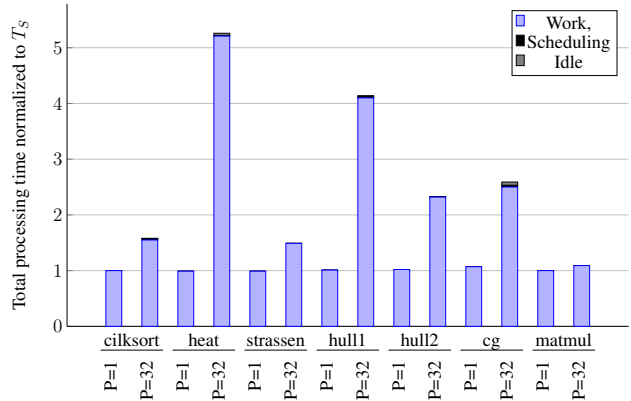

\paragraph{Work efficiency and work inflation of Cilk Plus.} The time a worker
spends can be categorized into three categories --- \defn{work time}, time
spent doing useful work (i.e., processing the computation), \defn{idle time},
time spent trying to steal but failing to find work to do, and
\defn{scheduling time}, time spent performing scheduling related tasks to
manage actual parallelism, such as frame promotions upon successful steals and
nontrivial syncs.  By looking at how much work time that all workers
collectively spend grows as the number of cores increases, one can gauge how
much work inflation impacts the scalability of the program.  On the other
hand, idle time is a good indication of how much parallelism a computation has
--- a computation that does not have sufficient parallelism is generally
marked by high idle time.  Finally, scheduling time indicates the runtime
overhead.  See \cite{AcarChRa17} for the formalization of this intuition.  

\figref{cilkplus} shows the normalized total processing times of six
benchmarks running on Cilk Plus, normalized to $\serial$, the execution time
of the corresponding serial elision.\footnote{The data is collected using the same
setup including input and base case sizes described in \secref{eval}.}  Note that 
\texttt{P=1} is simply $\work$, the time running on one worker, which includes the 
spawn overhead, but not scheduling or idle time, since no actual parallelism
is realized executing on one worker.

Cilk Plus has high work efficiency, as the ratio between $\work$ and $\serial$
is close to one.  Here, we did coarsen the base case --- when a
divide-and-conquer parallel algorithm reaches certain base case size, the code
stops spawning and calls the sequential version instead.  Coarsening helps
with spawn overhead and is a common practice for task-parallel code; it makes
a trade-off between spawn overhead and parallelism --- the smaller the base
case size the higher the parallelism and spawn overhead.  Typically one can
coarsen the base case size to mitigate spawn overhead but keeps it small
enough so that the computation has sufficient parallelism.  We chose the base
case sizes for these benchmarks by picking the ones that provide the best
$T_{32}$ raw execution times.  


For \texttt{P=32}, most benchmarks have very little scheduling time and idle time, 
indicating that the scheduler is efficient and there is sufficient parallelism
to saturate $32$ cores.  Finally, we look at work inflation.  Benchmarks
tested have work inflation ranging from $1.45\times$ to $5.24\times$, with the
exception of \texttt{matmul}.\footnote{The precise numbers for all benchmarks
are shown in \secref{eval}.}  The implementation of \texttt{matmul} is already
cache oblivious~\cite{FrigoLePr99} and thus had little work inflation to begin
with.  We use this benchmark as a baseline to show that, even for a benchmark
does not really benefit from NUMA-aware scheduling, the additional scheduling
mechanism in \NUMAWS does not adversely impact performance.  Moreover, the
data layout transformation helps when applied to \texttt{malmul}.  


\secput{numaws}{NUMA-Aware Task-Parallel Platform}



\NUMAWS extends Cilk Plus to incorporate mechanisms to mitigate NUMA effects.
An effective way to mitigate NUMA effects is to co-locate computations that
share data, and/or co-locate data and the computation that uses the data.  To
that end, \NUMAWS includes user-level APIs to provide locality hints,
extends the work stealing scheduler to be NUMA-aware, and schedules
computations according to the locality hints when possible.  In addition,
\NUMAWS also includes a simple data layout transformation that is applied to a
subset of benchmarks.  This section describes these extensions in detail.  

\subsecput{model}{Modifications to the Programming Model}

%
To facilitate NUMA awareness, the runtime at startup ensures that the
worker-\-thread-\-to-\-core affinity is fixed.  Since the programming model is
processor-oblivious, an application can run on any number of cores and sockets
within the constraints of the hardware resource.  We assume that the user
decides how many cores and how many sockets an application runs on when invoking
the application but does not change this configuration dynamically at runtime.

Given the user-specified number of cores and sockets, the runtime spreads 
out the worker threads evenly across the sockets and groups the threads on 
a given socket into a single group.  Each group forms a \defn{virtual 
place} that forms the basic unit for specifying locality.  

%
The locality API allows the user to query the number of virtual places in the
application code and specify which virtual place a spawned subcomputation
should ideally run on.  If the user specifies the locality for a spawned
subcomputation $G$, by default any computation subsequently spawned by $G$ is
also marked to have the same locality.  Such a default works well for
recursive divide-and-conquer algorithms, which task parallelism is well-suited
for.  For the benchmarks tested, this default means that only the top-level
root Cilk function needs to specify locality hints.  The API also includes
ways to unset or update the locality hints for a Cilk function.

\begin{figure}
\footnotesize
\begin{codebox}
\Procname{
//The $in$ array is a one-dimensional array with its physical memory}
\Procname{$\proc{MergeSortTop}$(int *$in$, int *$tmp$, int $n$)}
\li \If $n <$ BASE\_CASE
\li \Then $\proc{QuickSort}$($in$, $n$); \Comment{in-place sequential sort}
\li \Else
\li   \Comment{initialized p0, p1, p2, and p3 based on number of places}
\li   \Comment{$in$ and $tmp$ are partitioned appropriately.}
\li   cilk\_spawn $\proc{MergeSort}$($in$, $tmp$, $n/4$) \Comment{implicitly @p0}
\li   cilk\_spawn $\proc{MergeSort}$($in$ + $n/4$, $tmp$ + $n/4$, $n/4$); @p1
\li   cilk\_spawn $\proc{MergeSort}$($in$ + $n/2$, $tmp$ + $n/2$, $n/4$); @p2
\li               $\proc{MergeSort}$($in$ + $3n/4$, $tmp$ + $3n/4$, $n$ - $3n/4$); @p3
\li   cilk\_sync;
\li   int *$tmp1$ = $tmp$, *$tmp2$ = $tmp$ + $n/2$;
\li   cilk\_spawn $\proc{ParMerge}$($in$, $in$ + $n/4$, $n/4$, $n/4$, $tmp1$); @p0
\li   $\proc{ParMerge}$($in$ + $n/2$, $in$ + $3n/4$, $n/4$, $n$ - $3n/4$, $tmp2$); @p2
\li   cilk\_sync;
\li   $\proc{ParMerge}$($tmp1$, $tmp2$, $n/2$, $n$ - $n/2$, $in$); @ANY 
    \End 
\end{codebox}
\vspace{-2mm}
\caption{Pseudocode for the top-level function for parallel mergesort
with locality hints (as denoted by @p\# or @ANY).
The $\proc{MergeSort}$ is defined similarly as the
$\proc{MergeSortTop}$ but without the locality hint: it takes in an unsorted
input array, a temporary array, and their sizes, and sorts the input
array in place.  The $\proc{ParMerge}$ performs parallel merge; it takes in two 
sorted input arrays, their sizes, and an output array, and merges the two sorted 
inputs into the output.  Variables p0--p3 store IDs for virtual places based 
on the number of sockets used, but they do not have to be distinct (e.g., if 
less than four sockets are used).  The ANY indicates no place constraints 
and unsets the locality hint.} 
\label{fig:sort} \vspace{-4mm}
\end{figure}
  
\figref{sort} shows how one might use locality hints in a parallel 
mergesort, where $\proc{MergeSortTop}$ is the top-level root Cilk function:
it recursively sorts the four quarters of the input array in place, and
uses the temporary array to merge the sorted quarters back into the 
input array.  The locality hints are specified using the \code{@p#} notation,
where a \code{p#} is a variable storing the ID of a virtual place that the 
task (the corresponding function call and its subcomputation) should execute. 

Assuming the computation runs on four sockets, the code would initialize each
of the virtual places to the appropriate socket, and specify that the $i^{th}$
quarter should be sorted at the $i^{th}$ virtual place.  For merge, since we
have to merge two arrays sorted at two different places together, we simply 
specify its locality to be one of the virtual places that the inputs come 
from.  Even though we have only specified work for two virtual places for the
merge phase, the place specifications are only hints, and the \NUMAWS runtime  
will load balance work across all sockets dynamically as necessary.

\anote{elaborate on this} 
With continuation-stealing, the first spawned child is always executed by the
same worker who executes the corresponding \spawn, and thus we do not specify
a locality hint for the first spawn.  By default, the runtime always pins the
worker who started the root computation at the first core on the first socket
and thus implicitly the first spawned child always executes at the first
virtual place (at \code{p0} in the code).  If the user had specified a
locality hint for the first spawned child that differs from where the parent
is executing, the spawned child will obtain the user-specified locality, but
will not get moved to the specified socket immediately.  Rather, the
computation may get moved later as steals occur according to the lazy work
pushing mechanism described in \secref{runtime}.

In order to benefit from the locality hints, one should allocate the data 
on the same socket that the computation belongs to.  In this example, one
should allocate the physical pages mapped in the $i^{th}$ quarters of the 
$in$ and $tmp$ arrays from the socket corresponding to the $i^{th}$ virtual 
place.  We have developed library functions that allow the application code 
to do this easily at memory allocation time, but they are simply accomplished 
by calling the underlying \code{mmap} and \code{mbind} system calls that 
Linux OS provides.


The API described is an idealized API, which requires compiler support.  In
our current implementation, we manually hand compiled the application code by
explicitly invoking runtime functions to indicate the place specifications.
The transformation from the idealized API to the runtime calls is quite
mechanical and can easily be done by a compiler. 

\subsecput{runtime}{Modifications to the Scheduler}

The locality specified is only a hint, and the runtime tries to honor it 
with best effort and do so in a work-efficient fashion, incurring overhead
only on the span term.  Note that enforcing locality strictly
can impede runtime's ability to load balance, e.g., the computation contains
sufficient parallelism overall but most parallelism comes from computation
designated for a single socket.  Thus, the \NUMAWS runtime may execute the 
computation on a different socket against the locality hint if doing so
turns out to be necessary for load balancing.  

To obtain benefit using \NUMAWS, we expect the application code to
perform data partitioning with the appropriate locality hints specified.
However, because the runtime is designed to foremost treat load balancing
as the first priority and account for locality hints with best
effort, not specifying locality hints would not hurt performance much and 
result in comparable performance with one obtainable with the Cilk Plus  
runtime. 

At a high level, \NUMAWS extends the existing work-stealing scheduler with 
two NUMA-aware mechanisms:
\begin{closeitemize}
\item \textbf{Locality-biased steals:}
In the classic work stealing algorithm, when a thief steals, it chooses 
a victim uniformly at random, meaning that each worker gets picked in with
$1/P$ probability (and if the worker happens to pick itself, it tries again).
The locality-biased steals simply change the probability distribution of how a
worker steals, biasing it to preferentially steal work from victims running on
the (same) local socket over victims on the remote sockets.

\item \textbf{Lazy work pushing:}
The \defn{work pushing} refers to the operation that, a worker, upon receiving
a work item, instead of executing the work, pushes it to a different worker
to honor the locality hint.  Work pushing has been proposed in prior 
work (see \secref{related}).  However, the key distinction between our work
and prior work is that, \NUMAWS performs \defn{lazy work pushing}, in a way
that abides by the work-first principle and incurs overhead only on the span
term.  
\end{closeitemize}

\begin{figure}
\footnotesize
\begin{tabular}{|l|}
\hline
\begin{minipage}[t]{0.95\linewidth}
\vspace{\negspace}
\begin{codebox}
\Procname{$\id{F}$ executes sync:}
\li \If $\attrib{F}{stolen}$ = $\const{true}$ 
\li    \Then \Comment{$\id{F}$ must be a full frame and the deque is empty}
\li    $\id{success}$ = $\proc{checkSync}$(); \Comment{must do a nontrivial sync}
\li    \If $\id{success}$ \Then
\li       \If $\attrib{F}{place} \neq \attrib{worker}{place}$ \Then \label{li:syncpush1}
\li          \If $\proc{pushBack}$($\id{F}$) = $\const{true}$ \Then
\li             \Comment{another worker will resume $\id{F}$}
\li             $\id{next\_action}$ = $\const{STEAL}$
\li             return to scheduling loop
\li          \Else continue to execute $\id{F}$ 
             \End
\li       \Else continue to execute $\id{F}$ \label{li:syncpush2}
          \End
\li    \Else
\li       suspend $\id{F}$
\li       $\id{next\_action}$ = $\const{STEAL}$
\li       return to scheduling loop
       \End 
\li \Else continue to execute $\id{F}$ \Comment{nothing else needs to be done}
    \End
\end{codebox}
\end{minipage}%
\\ \hline
\begin{minipage}[t]{0.95\linewidth}
\vspace{\negspace}
\begin{codebox*}
\Procname{scheduling loop: 
\Comment{$\id{frame}$ is a either $\const{null}$ or the first root full frame}}
\li   \While $\id{computation-done}$ = $\const{false}$ \Do
\li      \If $\id{next\_action}$ = $\const{CHECK\_PARENT}$ \Then
\li         $\id{frame}$ = $\proc{checkParent}$();
\li         $\id{next\_action}$ = $\const{STEAL}$
                    \Comment{reset $\id{next\_action}$}
\li         \If $\id{frame} \neq \const{null}$ and \label{li:parentpush1}
               $\attrib{frame}{place} \neq \attrib{worker}{place}$ \Then
\li            \If $\proc{pushBack}$($\id{frame}$) = $\const{true}$ \Then
\li               \Comment{Another worker will resume $\id{frame}$}
\li               $\id{frame}$ = $\const{null}$
               \End
            \End                                   \label{li:parentpush2}
         \End
\li      \If $\id{frame}$ = $\const{null}$ \Then   \label{li:mailbox1}
\li         $\id{frame}$ = $\proc{popMailBox}$();  \label{li:mailbox2}
         \End
\li      \If $\id{frame}$ = $\const{null}$ \Then
\li         $\id{frame}$ = $\proc{biasedStealWithPush}$(); \label{li:biased-steals}
\li      \Else $\proc{resume}$($\id{frame}$) \End
\end{codebox*}
\end{minipage}%
\\ \hline
\end{tabular}
\caption{Modifications to the work-stealing scheduler in \NUMAWS. Here,
$\proc{pushBack}$() is the mechanism where the runtime pushes a full frame
to the virtual place that it has been designated for. $\attrib{F}{place}$ stores
the locality hint for $\id{F}$ and $\attrib{worker}{place}$ stores the 
virtual place that the worker belongs to.}
\label{fig:code-numaws}
\vspace{-4mm}
\end{figure}

\figref{code-numaws} shows \NUMAWS's modifications to work stealing to
incorporate these mechanisms, which we explain in detail next.  Compared to
the Cilk Plus scheduler, \NUMAWS may perform additional operations when
executing a nontrivial sync (\lirefs{syncpush1}{syncpush2}), inside the
scheduling loop (\lirefs{parentpush1}{mailbox2}), and it uses a modified
stealing protocol (\liref{biased-steals}).  The operations performed on a
spawn and a spawn returning are the same as shown in \figref{code-cilkplus},
so we do not repeat them here.

If the runtime performs work pushing indiscriminately, the overhead incurred
by work pushing would be on the work term instead of on the span
term.  Such overhead includes doing extra operations that do not advance
towards finishing the computation and synchronizing with the \defn{receiving}
worker (i.e., worker to push work to), which would result a work inefficient
runtime.

Instead in \NUMAWS, work pushing only occurs when a worker is handling a full
frame.  Recall from \secref{motivation} that, the data structures in the Cilk
Plus runtime are organized in accordance to the work-first principle: a frame
is either a shadow frame that has never been stolen before, or a full frame
that has been stolen successfully in the past and can contain actual
parallelism underneath (i.e., may have outstanding spawned children executing 
on different workers).  
Specifically, a worker only performs work-pushing in the following scenarios: 
\begin{closeitemize} 
\item A worker executes a non-trivial \sync successfully, and the synched full
frame is earmarked for a different socket via its locality hint
(\lirefs{syncpush1}{syncpush2}).

\item A worker returns from a spawned child, whose parent had executed
a nontrivial \sync unsuccessfully and therefore is suspended.  The
returning child is the last spawned child returning, and thus the parent is 
ready to be resumed at the continuation of \sync, but the parent is earmarked
for a different socket (\lirefs{parentpush1}{parentpush2}).

\item A thief steals successfully, and the stolen full frame is earmarked for
a different socket (part of $\proc{BiasedStealWithPush}$ in
\liref{biased-steals}).  \end{closeitemize}

Besides judiciously selecting certain control paths to perform work pushing,
another crucial aspect of lazy work pushing is how it selects the receiving
worker and how to push work without interrupting the receiving worker. 
This logic is implemented in $\proc{pushBack}$, which we explain below. 

When a worker needs to push a full frame to a designated socket, it
\emph{randomly} chooses a receiving worker on the designated socket to push
the frame to.  The chosen receiving worker may be busy working and likely have
a non-empty deque.  Thus, each worker besides managing a deque, also has a
\defn{mailbox}, that allows a different worker to deposit work designated for
the receiving worker without interrupting.  Crucially, the mailbox contains only
one entry; that is, each worker can have only one single outstanding ready
full frame that it is not actively working on.  A pushing worker may fail to
push work because the randomly chosen receiving worker is busy, and has a full
mailbox.  If the push fails, the pushing worker increments a counter on the
frame and tries again with a different randomly chosen receiving worker.  Once
the counter on the frame exceeds the \defn{pushing threshold} (a configurable
runtime parameter), the pushing worker simply takes the full frame and resumes
it itself.  We shall see in the analysis (\secref{analysis}) 1) how we can
amortize the cost of pushing against the span term, 2) why it's crucial to
have only a single entry for the mailbox, and 3) why there must be a
\emph{constant} pushing threshold.  

Besides potentially calling $\proc{pushBack}$, the steal protocol implemented
by $\proc{biasedStealWithPush}$ (\figref{code-numaws}: \liref{biased-steals}) 
differs from the original $\proc{randomSteal}$ (\figref{code-cilkplus}: 
\liref{steals}) in the following ways.  

First, the protocol implements locality biased steals.
Given the set of virtual places, the runtime configures the steal probability
distribution according to the distances between virtual places, where the
distances are determined by the output from \code{numactl}.  For instance, on
a 32-core 4-socket machine shown in \figref{numa}, each socket $i$ forms a
virtual place $i$.  When a worker on socket $0$ runs out of work, it will
preferentially select victims from the local socket (socket $0$) with the
highest probability, followed by victims from sockets that are one-hop away
(i.e., sockets $1$ and $2$) with medium probability, followed by victims from
the socket that is two-hop away (i.e., socket $3$) with the lowest
probability.

Second, since now a victim may potentially have a resumable frame in the
mailbox also (besides what's in the deque), a thief stealing needs to check
for both, but in a way that still retains the theoretical bounds.
Specifically, when a thief steals into a victim, it will flip a coin.  If the
coin comes up heads, it does the usual steal protocol by taking the frame at
the head of the deque (promoting it into a full frame).  If the coin comes up
tails, however, it checks the mailbox, followed by three possible outcomes: 1)
mailbox empty, and thus the thief falls back to stealing from the deque; 2)
mailbox is full, and the frame is earmarked for socket that the thief is on,
so the thief takes it; 3) mailbox is full, and the frame is earmarked for a
different socket; the thief then calls $\proc{pushBack}$ and performs the lazy
work pushing as described before until the frame reaches the pushing threshold
(in which case the thief can simply take it).

We shall see in the analysis (\secref{analysis}) 1) why the coin flip is
necessary and 2) why this biased steal protocol still provides provable
guarantees.

Finally, given that each worker now has a mailbox potentially containing work,
another small modification to the scheduling loop is needed.  When a worker
runs out of work and returns to the scheduling loop (in the
$\id{next\_action}$ = $\const{STEAL}$ case), it checks if something is in its
mailbox first (\figref{code-numaws}, \liref{mailbox2}).  If so, it simply
resumes it next.  Note that if a worker is back to the scheduling loop with
its $\id{next\_action}$ set to $\const{STEAL}$, its deque must be empty.

\subsecput{data-trans}{Simple Data Layout Transformation}   
An effective way to mitigate work inflation due to NUMA is to co-locate data
and computation that uses the data.  Even though modern OSs tend to provide
facility to bind pages to specific sockets, one must specify data allocation in
page granularity.  For applications that operate on 2D arrays such as various
matrix operations, the usual row-major order data layout is not conducive to
data and computation co-location, since most parallel algorithms use
divide-and-concur techniques which recursively subdivide the data into smaller
pieces that can span multiple rows but only part of the rows.  Thus, by the
time we reach the base case of a divide-and-conquer algorithm, the base case
would be accessing data that is scattered across multiple physical pages,
making it challenging to co-locate data and the computation. 

\begin{figure}
\begin{minipage}[b]{0.48\linewidth}
\centering
\includegraphics[width=1.4in]{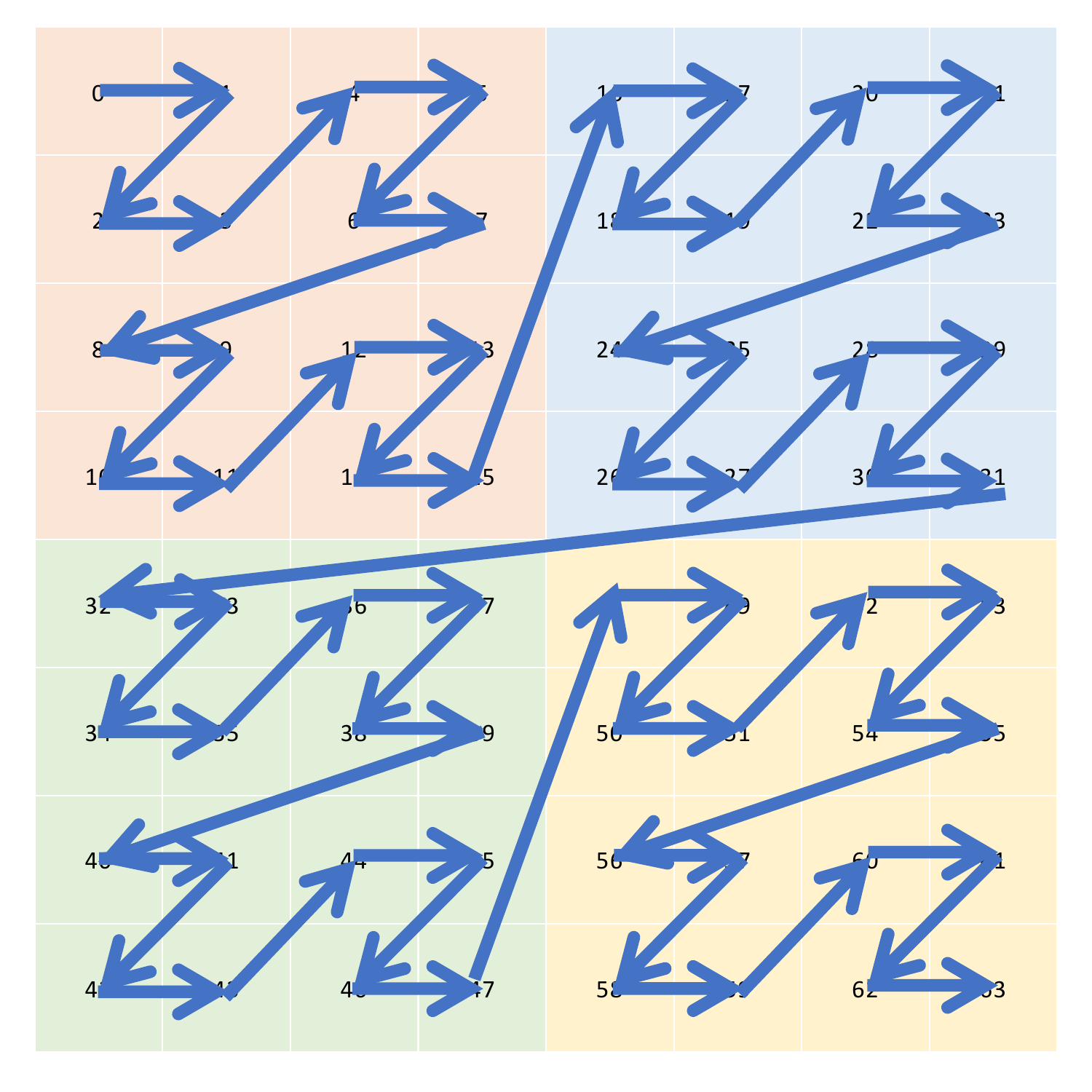}\\
\textbf{(a) Z-Morton}
\end{minipage}
\begin{minipage}[b]{0.48\linewidth}
\centering
\includegraphics[width=1.4in]{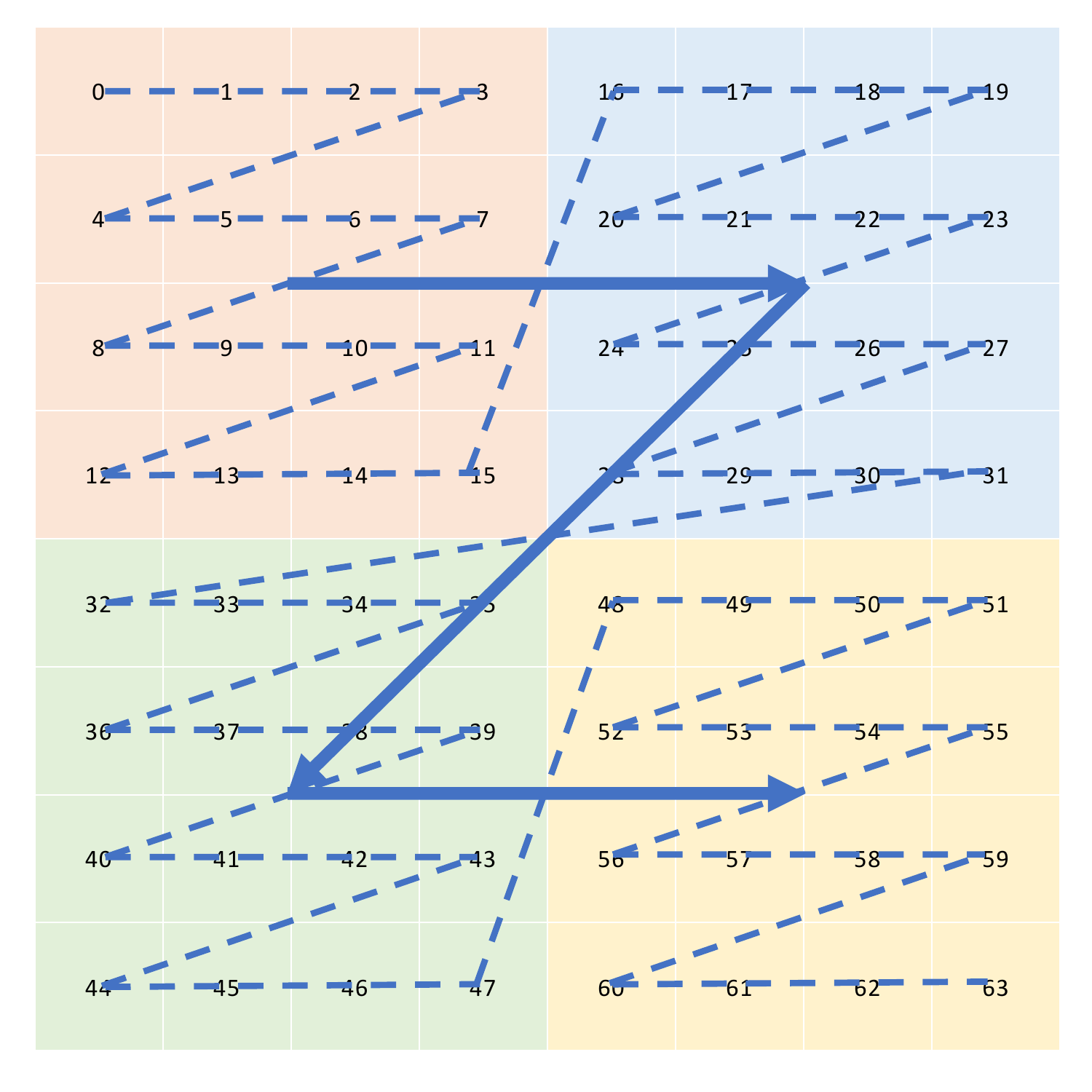}\\
\textbf{(b) blocked Z-Morton}
\end{minipage}
\caption{\textbf{(a)} Ordinary Z-Morton layout on the cell-by-cell basis
where every element lies on a recursive Z curve.
\textbf{(b)} The blocked Z-Morton layout used in our data layout 
transformation, where the blocks are laid out recursively on the Z curve, 
and data within each block is laid out in row-major order.}
\label{fig:morton}
\vspace{-4mm}
\end{figure}

One way to fix it is to use \defn{Z-Morton} layout (also called
\defn{cache-oblivious bit-interleaved} layout), which interleaves the bits
when calculating the index so that the data is laid out in a
recursive Z curve~\cite{FrigoLePr99} (\figref{morton}a).  
Computing indices for Z-Morton layout on the cell-by-cell basis is costly, however.  
\anote{Get better citation for Z-morton}

To achieve co-location, our platform provides APIs to allow the user to
perform data layout transformation.  The data layout transformation simply
lays out blocks in Z-Morton layout and lays out the data within each block in
row-major order, as shown in \figref{morton}b.  Doing so has the following
benefits: 1) the data accessed by the base case of a divide-and-conquer
algorithm that utilizes 2D arrays will be contiguous in memory; and 2) since
the bit interleaving needs to be computed only for the block indices, we save
on overhead for index computation.  While the use of Z-Morton layout is not
new, in \secref{eval}, we show that the blocked Z-Morton layout works well
with existing NUMA systems.

\secput{analysis}{\NUMAWS's Theoretical Guarantees}

\NUMAWS provides the same execution time and cache miss bounds as classic work
stealing.  Specifically, \NUMAWS executes a computation with $\work$ work and
$\spa$ span in expected time $\work / P + O(\spa)$ on $P$ cores, and the
number of successful steals can be bounded by $O(P\spa)$, although the
constant in front of the $\spa$ term would be larger than that for classic
work stealing.  The bound on the number of steals allows one to constrain the
additional cache misses with respect to sequential execution for private
caches~\cite{AcarBlBl00}.  The structure of the theoretical analysis largely
follows that described by Arora et.\@ al ~\cite{AroraBlPl98} but with some
delta.  We provide the high level intuitions and describe the delta here. 

We first explain the intuition behind the original work stealing analysis
using the potential function formulation by Arora et. al~\cite{AroraBlPl98},
henceforth referred to as the ABP analysis.  Here, we assume a dedicated 
execution environment.\footnote{The ABP analysis provides bounds for both 
dedicated and multiprogrammed environments.}

In the ABP analysis, a computation (a program for a given input) is modeled as
a \defn{directed acyclic graph (dag)}, where each node represents a
\defn{strand}, a sequence of instructions that contain no parallel control
that must be executed by a single worker, and each edge represents a
dependence --- if there is an edge between node $u$ and $v$, then $v$ cannot
execute until $u$ finishes.  Based on this model, a \spawn generates a node
with two out-degree, one to the spawned child and one to the
continuation; a \sync generates a node with multiple in-degree, one from each
spawned child joined by the \sync.  Assuming each node takes a unit time step
to compute (if a strand takes more than one time step to compute, it is split
into a chain of nodes), the work is then defined as the total number of nodes
in the dag, and span is the number of nodes along a longest path in the dag.  

At each time step, a worker is either doing useful work, i.e., executing a
node (call it a \defn{work step}), or stealing including failed steal attempts
(call it a \defn{steal step}).  The total number of work steps across all
workers is bounded by $\work$, because if the workers have collectively spent
$\work$ time steps doing useful work, the computation would be done.  Thus,
the key is to bound the number of steal steps.  As long as we can bound the
number of steal steps to $O(P\spa)$, the bound follows, since the execution
time on $P$ workers is the number of total steps divided by $P$.

Informally, the intuition of the analysis is as follows.  The non-empty deques
contain nodes ready to be executed; in particular, there is some node that is
``critical'' --- a node on the span and if executed the span will decrease by
one.  Due to how work stealing operates, such a critical node must be at the
head of some deque.  Thus, after $O(P)$ steal attempts (since each deque has
$1/P$ probability to be stolen from), the critical node is likely to be
executed.  Thus, after $O(P\spa)$ steal attempts, we exhaust the span, thereby
bounding the steal steps.

To show this formally, the ABP analysis uses a potential function formulation
to bound the total number of steal steps.  The nodes pushed onto a worker's
deque are all ready to be executed and has certain amount of potential,
defined as a function of both the span of the computation and the ``depth'' of
the node in the dag --- roughly speaking, one can think of the potential
function as, ``how far away the node is from the end of the computation.'' Due
to the way work-stealing operates, it maintains a property called the 
\defn{top-heavy deques}\footnote{In the ABP analysis~\cite{AroraBlPl98}, they
refer to the head of the deque as its ``top'' and the tail as its ``bottom,''
and hence the name of the lemma.} (Lemma $6$ in \cite{AroraBlPl98}): the node
at the head of a non-empty deque constitutes a constant fraction of the
overall potential of the deque.  Moreover, after $O(P)$ steal attempts, with
constant probability, the overall potential of the computation decreases by a
constant fraction (Lemmas $7$ and $8$ in \cite{AroraBlPl98}).  The intuition
is that, after that many steals, the critical node at the head of some deque
are likely to get stolen and executed, thereby decreasing the potential by a
constant fraction.  Since the computation started out with certain amount of
potential, defined in terms of the span of the computation, there cannot be
more than $O(P\spa)$ steal attempts before the potential reaches $0$ (i.e.,
the computation ends).

There are two key elements in this argument.  First, if a deque is non-empty,
the head of a deque contains a constant fraction of the potential in the
deque.  Second, after $O(P)$ steal attempts, the critical node at the head of
some deque gets executed.

The property of top-heavy deques still holds in \NUMAWS, because a worker
still pushes to the tail and a thief steals form the head.  In \NUMAWS,
however, 1) the steal probability has changed (no longer uniformly at random);
2) a thief randomly chooses between the victim's mailbox and its deque; 3)
even if a thief steals successfully, it can push the stolen frame back to the
designated socket (into a mailbox) instead of executing it immediately.  We
first argue why these changes do not jeopardize the analysis.  Then we show
how we bound the additional cost of work pushing.

It turns out that, as long as the critical node gets stolen and executed with
probability $1/(cP)$ for some non-zero constant $c$, we can show that $O(P)$
steal attempts cause the overall potential to decrease by a constant fraction,
which in turn bounds the number of steals to $O(P\spa)$.  In \NUMAWS, the
critical node could either be at the head of some deque or in some worker's
mailbox (via lazy work pushing).  

Specifically, the following lemma is a straightforward generalization of
Lemmas $7$ and $8$ in \cite{AroraBlPl98}:
\begin{lemma}
Let $\Phi(t)$ denote the overall potential of the computation at time
step $t$.  Assuming the probability of each deque being the target of a steal
attempt is at least $1/X$, then after $X$ steal attempts, the potential is at
most $\Phi(t)/4$ is at least $1/4$.
\end{lemma}
Here, let $1/X$ be $1/(2cP)$ for some constant $c > 0$ that corresponds to the
probability of stealing into a worker on the most remote socket.  The factor
of $2$ is due to the fact that a thief only steals into a victim's deque with
$1/2$ probability once a given victim is chosen (the coin flip that decides
whether to steal into the deque or the mailbox).

The rest of the proof in the ABP analysis follows, and one can show that the
number of steal steps are bounded by $O(P\spa)$ (shown in Theorem $9$ in 
\cite{AroraBlPl98}).  Naturally, the smaller the probability of
stealing into the critical node, the larger the constant hidden in the
$O(P\spa)$ term, and thus \NUMAWS has a larger constant hidden in front of the
$\spa$ term. 

It is important that a thief flips a coin to decide whether it should steal
into a victim's mailbox versus its deque instead of always looking into the
mailbox first.  The coin flip guarantees that the critical node is stolen with
probability at least $1/(2cP)$. If the thief always looks into the mailbox,
the critical strand could be at the head of some deque and never gets stolen.
Having a mailbox of size one also guarantees that, since there can be only one
entry in the mailbox at any given point.  A constant-sized mailbox can work,
but would complicate the argument and require imposing an ordering on how the
mailbox is accessed --- a mailbox with multiple entries would need to maintain
the top-heavy deques property as well.  

Now we consider the extra overhead incurred by work pushing.  Since pushing
work does not directly contribute to advancing the computation, we need to
bound the time steps workers spent pushing work.  We will bound the cost of
pushing by amortizing it against successful steals and show that only a
constant number of pushes can occur per successful steal.  

A worker performs work pushing only on full frames that belong to a different
socket under these scenarios: a successful non-trivial sync, last spawned
child returning to a suspended parent, and a successful steal.  For each of
such events, one can attribute the event to some successful steal occurred
that led to the event.  Moreover, there can be at most two such events counted
towards a given successful steal, since a frame only performs a nontrivial
sync if it has been stolen since its last successful sync.  Since only at most
two events can be counted towards a successful steal, and only at most
constant number of pushes can occur per event due to the pushing threshold
(defined in \secref{numaws}), we can amortize the cost of pushing against
successful steals and upper bound that by $O(P\spa)$ as well.  This
amortization argument utilizes the fact that we have a constant pushing
threshold and a single-entry mailbox.  

Finally, in classic work stealing, the number of additional cache misses on
private caches due to parallel execution is simply bounded by the number of
successful steals~\cite{AcarBlBl00} --- $O(P\spa C)$ for a private cache of
size $C$.  The intuition is that, each successful steal forces the worker to
refill its private cache.  In \NUMAWS, the same bound follows since both the
steals and work pushing are bounded by $O(P\spa)$.

\secput{eval}{Empirical Evaluation}

This section empirically evaluates \NUMAWS.  \NUMAWS has similar work
efficiency and low scheduling overhead as in Cilk Plus, but it mitigates work
inflation and thus provides better scalability.  Moreover, \NUMAWS maintains
the same processor-oblivious model, and all benchmarks tested scale as the
number of cores used increases.

\paragraph{Experimental setup.} 
We ran all our experiments on a $32$-core machine with 2.20-GHz cores on four
sockets (Intel Xeon E5-4620) with the same configuration shown
in~\figref{numa}.  Each core has a 32-KByte L1 data cache, 32-KByte L1
instruction cache, and a 256-KByte L2 cache.  Each socket shares a 16-MByte
L3-cache, and the overall size of DRAM is 512 GByte.  All benchmarks are
compiled with Tapir~\cite{SchardlMoLe17}, a LLVM/Clang based Cilk Plus
compiler, with \code{-O3} running on Linux kernel version 3.10 with NUMA
support enabled.  Each data point is the average of $10$ runs with standard
deviation less than $5\%$.  When running the vanilla Cilk Plus, we tried both
the first-touch and interleave NUMA policies for each benchmark and used the
configuration that led to the best results.

\paragraph{Benchmarks.} Benchmark \code{cg} implements conjugate gradient that
solves system of linear equations in the form of $Ax = b$ with a sparse input
matrix $A$.  Benchmark \code{cilksort} performs parallel mergesort with
parallel merge.  Benchmark \code{heat} implements the Jacobi-style heat
diffusion on a 2D plane over a series of time steps.  Benchmark \code{hull}
implements quickhull to compute convex hull.  The algorithm works by
repeatedly dividing up the space, drawing maximum triangles, and eliminating
points inside the triangles.  When there are no more points outside of the
triangles, we have found the convex hull.  Since \code{hull}'s work and span
can differ greatly depending on the input data points, we ran it on two
different data sets: one with randomly generated points that lie within a
sphere (\code{hull1}), and another with randomly generated points that lie on
a sphere (\code{hull2}).  There is a lot more computation in \code{hull2},
since the algorithm has a harder time eliminating points.  
Benchmark \code{matmul} implements a eight-way divide-and-conquer matrix
multiplication with no temporary matrices.  Benchmark \code{strassen}
implements a matrix multiplication algorithm that performs seven recursive
matrix multiplications and a bunch of additions.  Most benchmarks are
originally released with MIT Cilk-5~\cite{FrigoLeRa98}, except for \code{cg},
which comes from the NAS parallel benchmarks~\cite{BaileyBaBa91} and
\code{hull}, which comes from the problem-based benchmark
suite~\cite{ShunBlFi12}.  Two benchmarks benefit from the data layout
transformation (\secref{data-trans}): \code{matmul} and \code{strassen}, and
we also ran the versions with the data layout transformation (\code{matmul-z}
and \code{strassen-z}) on both platforms.  For a given application, we used
the same input and base case sizes for both platforms.  

\subsection{Work Efficiency and Scalability} 

\begin{figure*}[t]
\centering
\begin{footnotesize}
\newcommand{\oh}[1]{\scriptsize (#1$\times$)}
\newcommand{\mch}[3]{\multicolumn{#1}{#2}{\textit{#3}}}
\begin{tabular}{ccr|
    r@{\hspace{0.5\tabcolsep}}rr@{\hspace{0.5\tabcolsep}}r|
    r@{\hspace{0.5\tabcolsep}}rr@{\hspace{0.5\tabcolsep}}r}
    & \textit{input size /}   & \mch{1}{c|}{serial} 
    & \mch{4}{c|}{Cilk Plus} & \mch{4}{c}{\NUMAWS} \\
\textit{benchmark} & \textit{base case size}
    & \mch{1}{c|}{$\serial$} & \mch{2}{c}{$\work$} & \mch{2}{c|}{$T_{32}$} 
    & \mch{2}{c}{$\work$} & \mch{2}{c}{$T_{32}$} \\
\hline
\texttt{cg} & $75k \times 75/n/a$ 
& 360.00 & 385.41 & \oh{1.07} &  29.39 & \oh{13.11} & 384.48 & \oh{1.07} & 14.89 & \oh{25.82}\\
\texttt{cilksort} & $1.3e8 / 1k$ 
& 20.38 & 20.47 & \oh{1.00} &  0.96 & \oh{21.28} & 20.95 & \oh{1.03} & 0.79 & \oh{26.58}\\
\texttt{heat} & $16k \times 16k \times 100 / 16k \times 10$ 
& 83.48 & 83.05 & \oh{0.99} &  13.78 & \oh{6.03} & 83.05 & \oh{0.99} & 5.95 & \oh{13.97}\\
\texttt{hull1} & $100000k / 10k$ 
& 4.08 & 4.12 & \oh{1.01} & 0.53 & \oh{7.71} & 4.11 & \oh{1.01} & 0.45 & \oh{9.04}\\
\texttt{hull2} & $100000k / 10k$ 
& 44.22 & 44.95 & \oh{1.02} &  3.29 & \oh{13.67} & 44.69 & \oh{1.01} & 2.09 & \oh{21.34}\\
\texttt{matmul} & $4k \times 4k / 32 \times 32$ 
& 190.86 & 191.03 & \oh{1.00} &  6.45 & \oh{29.60} & 190.39 & \oh{1.00} & 6.40 & \oh{29.76}\\
\texttt{matmul-z} & $4k \times 4k / 32 \times 32$ 
& 73.63 & 73.64 & \oh{1.00} &  2.34 & \oh{31.44} & 73.65 & \oh{1.00} & 2.35 & \oh{31.29}\\
\texttt{strassen} & $8k \times 8k / 16 \times 16$ 
& 112.82 & 111.78 & \oh{0.99} &  5.08 & \oh{22.00} & 111.99 & \oh{0.99} & 5.01 & \oh{22.37}\\
\texttt{strassen-z} & $8k \times 8k / 16 \times 16$ 
& 80.43 & 82.03 & \oh{1.02} &  3.46 & \oh{23.69} & 81.78 & \oh{1.02} & 3.47 & \oh{23.59}\\
\end{tabular}
\caption{The execution times in seconds for the benchmarks: its serial elision, running 
on Cilk Plus, and on \NUMAWS.  The data layout transformation is applied to two
benchmarks: \texttt{matmul} and \texttt{strassen}, denoted as \texttt{matmul-z} and
\texttt{strassen-z}.  The numbers in parentheses under the $\work$ columns 
indicate spawn overhead, (i.e., $\work$/$\serial$).  The numbers in parentheses 
under the $T_{32}$ columns show scalability (i.e., $\work / T_{32}$).}
\label{fig:overview}
\end{footnotesize}
\vspace{-3mm}
\end{figure*}

We first provide the overview of
our results.  \figref{overview} shows the $\serial$, $\work$, and $T_{32}$
executing on Cilk Plus and on \NUMAWS.  As expected, the $\work$ for all
benchmarks are similar for the two platforms, since the code are similar with
the exception of linking with different scheduler.  We compute the spawn
overhead by dividing $\work$ with the corresponding $\serial$ (shown in
parentheses under $T_1$).  Like in Cilk Plus, with appropriate coarsening,
\NUMAWS retains high work efficiency.  \NUMAWS does not incur any additional
overhead on the work term to achieve NUMA awareness.  For $T_{32}$, \NUMAWS
was able to achieve better scalability compared to Cilk Plus   for most
benchmarks (shown in parentheses under $T_{32}$).

Both \code{matmul} and \code{strassen} benefit from the data layout
transformation (i.e., comparing with the \code{-z} version). The blocked Z
layout helps when used in matrix multiplications, because the index calculation
incurs little additional overhead, and it traverses the matrices in a way that
enables the prefetcher.  

Beyond data layout transformation, \NUMAWS does not provide more benefit,
which is expected, because \code{matmul} readily obtains good scalability to
begin with, and we didn't use locality hints in \code{strassen}.  It's
challenging to specify sensible locality hints in \code{strassen} due to how
the algorithm works.  Sub-matrices of the inputs are used in different parts
of the computation, and thus the data necessarily has to be accessed by
multiple sockets.  One easy way to specify locality hint for \code{strassen}
is to perform a eight-way divide-and-conquer matrix multiplication at the
top-level, and only perform the seven-way divide starting from the second
level of recursion.  Doing so would allow one to specify locality hint at the
top level.  We attempted that strategy, but it turns out that, the $T_{32}$
performance of the top-eight-way version is comparable to the version reported
with no locality hint.  This is because the top-eight-way version indeed have
less work inflation, but at the expense of 15\% increases in overall $\work$,
because we are not getting the $O(n^{\lg 7})$ work at the top level.  Thus,
all things considered, we chose this version over the top-eight-way, since it
is more work efficient, and provides comparable $T_{32}$ execution time.

\subsection{Scheduling Overhead and Work Inflation}

\begin{figure*}[t]
\centering
\begin{footnotesize}
\newcommand{\oh}[1]{\scriptsize (#1$\times$)}
\newcommand{\mch}[3]{\multicolumn{#1}{#2}{\textit{#3}}}
\begin{tabular}{c|rr@{\hspace{0.5\tabcolsep}}rrr|rr@{\hspace{0.5\tabcolsep}}rrr}
     & \mch{5}{c|}{Cilk Plus} & \mch{5}{c}{\NUMAWS} \\ 
\textit{benchmark}
    & \mch{1}{c}{$T_{1}$}  & \mch{2}{c}{$W_{32}$} & \mch{1}{c}{$S_{32}$} & \mch{1}{c|}{$I_{32}$}
    & \mch{1}{c}{$T_{1}$}  & \mch{2}{c}{$W_{32}$} & \mch{1}{c}{$S_{32}$} & \mch{1}{c}{$I_{32}$} \\ 
\hline
\texttt{cg} 
& 385.41  &  898.60 & \oh{2.33} &  10.44 & 21.48  & 384.48 & 443.63 & \oh{1.21} & 6.75 &  17.36  \\
\texttt{cilksort} 
&  20.47  &   31.56 & \oh{1.54} &  0.33  &  0.14  & 20.95 & 25.39 & \oh{1.21}&  0.15  & 0.06 \\
\texttt{heat} 
&  83.05  &  435.11 & \oh{5.24} &  1.09 &  2.96  & 83.05 & 186.83 & \oh{2.25} &  0.43  &  1.89  \\
\texttt{hull1} 
&   4.12  &  16.71 & \oh{4.05}  &  0.08 &  0.09 &  4.11 & 14.50 & \oh{3.53} &  0.05 &  0.14 \\
\texttt{hull2} 
& 44.95  &  102.45 & \oh{2.28} &  0.17  &  0.23  & 44.69 & 69.62 & \oh{1.56} &  0.16 &  0.28 \\
\texttt{matmul} 
&  191.03 &  207.78 & \oh{1.09} &  0.25  &  0.15  & 190.39 & 202.79 & \oh{1.07} &  0.40 &  0.52\\
\texttt{matmul-z} 
&  73.64 &  74.99 & \oh{1.02} &  0.25  &  0.11  & 73.65 & 74.79 & \oh{1.02} &  0.28 &  0.29\\
\texttt{strassen} 
&  111.78 &  168.17 & \oh{1.50} &  0.39  &  0.52  & 111.99 & 168.28 & \oh{1.50} &  1.22 &  0.39\\
\texttt{strassen-z} 
&  82.03 &  119.44 & \oh{1.46} &  0.88  &  0.75  & 81.78 & 118.34 & \oh{1.45} &  2.29 &  0.43\\
\end{tabular}
\end{footnotesize}
\caption{$\work$ shows the one-core running time on each platform.
$W_{32}$, $S_{32}$, and $I_{32}$ show the work time, scheduling time, and
idle time, respectively, when running on $32$ cores.  The numbers in parentheses 
next to $W_{32}$ indicates the work inflation (i.e., $W_{32}/\work$) compared
with the $\work$ from the same platform.}
\label{fig:work-inf}
\vspace{-3mm}
\end{figure*}

We examine the detailed breakdown of $T_{32}$ next.  \figref{work-inf} shows
the work time running on one core, and the work, scheduling, and idle times
running on $32$ cores for both platforms.  We note that, like in Cilk Plus,
\NUMAWS has little scheduling overhead.  Since the scheduling overhead is
already low in Cilk Plus to begin with, the improved scalability of \NUMAWS
largely comes from mitigated work inflation.  

By comparing $W_{32}$ (work time on $32$ cores) with its respective $\work$,
one can gauge the work inflation for both platforms.  Note that one cannot
entirely avoid work inflation, since any shared data (i.e., multiple workers
taking turns to write to the same memory location) or any work migration
(i.e., successful steals) will inevitably incur work inflation.  However,
compared to Cilk Plus, \NUMAWS indeed mitigates work inflation. 

In general, \code{cg}, \code{cilksort}, \code{heat}, and \code{hull} (both
inputs) obtain visible decrease in work inflation when running on \NUMAWS;
\code{matmul} has little to begin with, and \code{strassen} did not use
locality hints as explained in earlier.  Benchmark \code{hull1} has a higher
work inflation than \code{hull2}, because for the particular input used for
\code{hull1}, points were eliminated quickly.  Thus, the majority of the
computation time is spent doing parallel prefix sum, where the forward and
backward propagations touch different parts of the data array and simply does
not have much locality.

\subsection{Maintaining Processor-Oblivious Model}

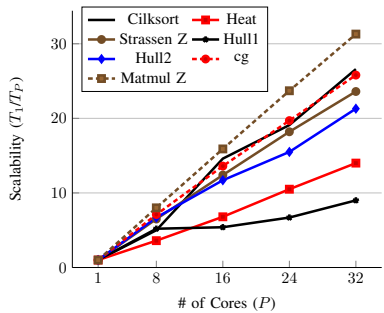
\begin{figure}
\begin{minipage}{0.63\columnwidth}
\begin{tikzpicture}[scale=0.60]
\pgfplotstableread{
X	cg	cilksort	heat	hull1	hull2 	lu	strassen	strassenz	fft	matmul	matmulz
1.0	1.0	1.0	1.0	1.0	1.0	1.0	1.0	1.0	1.0	1.0	1.0
8.0	7.1	5.0	3.6	5.2	6.7	8.0	6.1	6.5	6.6	7.6	8.0
16.0	13.6	14.6	6.8	5.4	11.7	15.4	11.6	12.4	8.2	15.2	15.9
24.0	19.7	19.1	10.5	6.7	15.5	22.3	16.9	18.2	8.8	22.7	23.7
32.0	25.8	26.6	14.0	9.0	21.3	29.8	22.4	23.6	16.4	29.8	31.3
}\datatable

\begin{axis}[
    axis lines*=left, ymajorgrids,
    every axis plot/.append style={ultra thick},
    xlabel=\# of Cores ($P$),
    ylabel=Scalability ($T_1$/$T_P$),
    legend columns=2,
    ymin=0,
    xtick=data,
    legend style={at={(0.03,0.85)},anchor=west}
]

\addplot [width=5mm] table [x=X,y=cilksort]{\datatable};\addlegendentry{Cilksort}
\addplot table [x=X,y=heat]{\datatable};\addlegendentry{Heat}
\addplot table [x=X,y=strassenz]{\datatable};\addlegendentry{Strassen Z}
\addplot table [x=X,y=hull1]{\datatable};\addlegendentry{Hull1}
\addplot table [x=X,y=hull2]{\datatable};\addlegendentry{Hull2}
\addplot table [x=X,y=cg]{\datatable};\addlegendentry{cg}
\addplot table [x=X,y=matmulz]{\datatable};\addlegendentry{Matmul Z}
\end{axis}
\end{tikzpicture}
\end{minipage}
\hspace{-3mm}
\begin{minipage}{0.35\columnwidth}
\vspace{-2mm}
\caption{The scalability of benchmarks running on \NUMAWS.  The x-axis shows
$P$, the number of cores used, and the y-axis shows $\work$/$T_P$, the
scalability.  Threads are packed onto sockets tightly and the smallest
number of sockets is used, i.e., for $24$ cores, 3 sockets are used.}
\label{fig:scalability}
\end{minipage}
\vspace{-4mm}
\end{figure}

Next, we show that, \NUMAWS retains the processor-oblivious programming model
and the same benchmarks work well across different number of cores.
\figref{scalability} shows the scalability ($\work/T_P$) plot across
benchmarks.  The workers are packed tightly using the smallest number of
sockets.  For a given benchmark, its data points are collected using the same
user applications without modification.  The only thing that changed is the
input argument to the runtime specifying the number of cores and sockets to
use.  Even though we ran the program on different number of sockets, there is
no need to change the program.  The program queries the runtime how many
physical sockets are used during the initialization phase and initializes
variables storing the IDs of virtual places accordingly depending on how many
sockets are used.  

As can be seen in \figref{scalability} the scalability curves are smooth,
indicating that the application indeed gains speedup steadily as we increase
the number of cores.  An exception is \texttt{hull1}, which as explained
earlier, spends majority of the computation time on prefix sum, which does not
have much locality.  Its scalability clearly degrades once we move from a
single socket to multiple sockets.

\secput{related}{Related Work}

Researchers have observed the problems of scaling task parallel programs due
to work inflation, and proposed various mechanisms and scheduling policies to
mitigate the effect~\cite{GuoZhCa10,ChenHuGu11,LifflanderKrKa14,OlivierPoWh12,
ChenGuHu12,OlivierDeSuSc12,MuddukrishnaJoVl13,SimhadriBlFi14,DrebesHeDr14,
DrebesPoHe16,MajoGr17}.  None of the proposed platforms achieve all the desired
goals, however.  In particular, none of them focuses on achieving work
efficiency nor provides provably efficient time bound. 

One common approach is to utilize some kind of work-stealing hierarchy, where
the scheduler employs a centralized shared queue among workers within a socket
so as to load balance across sockets.  Work described by~\cite{ChenHuGu11,
OlivierDeSuSc12, OlivierPoWh12, ChenGuHu12, ChenGuGu14, DrebesHeDr14,
DrebesPoHe16} take such an approach.  In order to aid load balancing across
sockets, work described by~\cite{ChenHuGu11, ChenGuHu12, ChenGuGu14} also
propose heuristics to perform the initial partitioning of the work so that
work distributed to different sockets can be somewhat balanced to begin with.
It's not clear the scheduler proposed by these prior work provide the same
theoretical guarantees as the classic work stealing.  Moreover, if the initial
partitioning was not done in a load-balanced way, the performance may  
degrade, since load balancing among sockets require more centralized control.
In contrast, our work still utilizes randomized work stealing and provides
sound guarantees.  

Besides having a hierarchy for work stealing, prior work~\cite{DrebesHeDr14,
DrebesPoHe16} also explored the notion of work pushing, but the key
distinction between these work and ours is that, \NUMAWS performs work pushing
in a work-efficient way.  Without judicious selection of when to performing
work pushing, the scheduler can incur high pushing overhead on the
work term, causing the parallel overhead ($\work$/$\serial$) to increase.

Related to work-pushing, the notion of mailbox is first proposed
by~\cite{AcarBlBl00}.  In their paper, Acar et.\@ al provided the analysis for
how to bound cache misses for private caches for the classic work stealing 
algorithm.  The paper also proposed a heuristic for obtaining better locality
for iterative data parallel code, where the program iteratively executes a
sequence of data-parallel loops that access the same set of data over and
over.   The assumption is that, the program can obtain better locality if the 
runtime can keep the same set of data-parallel iterations on the same worker.
Upon a spawn, if the work item being pushed onto the deque has a different affinity 
(as determined by the data-parallel loop indices) from the executing worker, 
the work item is pushed onto both the executing worker's deque and the designated 
worker's mailbox, a FIFO queue with multiple entries.  
The use of mailbox is proposed as a heuristic, and their
analysis does not extend to include the heuristic.  

The use of mailbox is subsequently incorporated into Intel Threading Building
Blocks (TBB)~\cite{Reinders07,IntelTBBManual}.  Majo and Gross extended TBB
to be NUMA-aware in a portable fashion.  Our proposed programming API took inspiration 
from them.  However, their work has similar downside that, it's not work efficient 
and does not provide provable guarantee.  

Work by~\cite{GuoZhCa10} also utilizes the notion of places, and the
programmer can specify that a spawned task being executed at a specific place.
The scheduler restricts such tasks to be executed exactly at the designated
place, however, which can impede scheduler's ability to load balance and thus
leads to inefficient execution time bound.

Work by~\cite{LifflanderKrKa13} targets specifically workload that performs
the same parallel computation over and over and thus their mechanism is
designed specifically for the set of programs that exhibit such behaviors.  In
their work, the workers records the steal pattern during the first iteration,
and replay the same steal pattern in subsequent iterations, thereby gaining
locality.

Work by~\cite{BlellochFiGi11, SimhadriBlFi14} studies a space-bounded
scheduler, which provides provable guarantees for bounding cache misses for
shared caches, but may sacrifice load balance as a result. 

Work by~\cite{MaglalangKrAg17} proposes a locality-ware task graph scheduling
framework, which provides a provably good execution time bound.  However, the 
framework is designed for task graph computations which has a different 
programming model.

Finally, work by~\cite{SuksompongLeSc18} proposed a locality-aware scheduler 
called \defn{localized work stealing}.  Give a computation with well-defined 
affinities for work items, each worker keeps a list of other workers who might 
be working on items belonging to it.  That is, whenever a thief steals a work 
item, it will check the affinity of the work item and add its name onto the 
owner's list (who might be a different worker from the victim).  Whenever a worker 
runs out of work, it checks the list first and randomly select a worker from 
the list to steal work back.  
This steal-back mechanism, like the lazy work pushing in our work, can be 
amortized against steals.  However, since a worker is required to check the 
list when it runs out of work to do 
their bound is slightly worse.  The work is primarily theoretical and has not 
been implemented.   
       
\secput{concl}{Conclusion and Future Direction}

In this paper, we have shown that \NUMAWS, like the classic work stealing,
provides strong theoretical guarantees.  Moreover, its implementation 
is work efficient and can provide better scalability by mitigating work 
inflation.  We conclude by discussing some of its limitations and potential
future directions.  \NUMAWS is designed to give the programmer a finer control
over where a task is executed.  In order to mitigate work inflation, however,
the programmer still needs to provide locality hints and allocates /
partitions data in such a way that allows the task and its data to be
co-located.  First, the programmer needs to use the runtime to query the
number of sockets and perform the appropriate data partitioning, and thus
cannot be entirely socket oblivious.  Second, it may be challenging to
partition data and provide sensible locality hints for algorithms that perform
random memory accesses (i.e., a task may access data scattered across
sockets).  Examples of such algorithms include various graph algorithms, or
\code{strassen} we tested.  While \NUMAWS won't degrade the performance of
such algorithms, it won't bring benefit, either.  Interesting future
directions include devising a programming interface that allows the programmer
to be socket oblivious and investigating how one may mitigate NUMA effects in
algorithms where it's challenging to co-locate a task and its data.

\section*{Acknowledgments}

We thank Suyeon Kang and Ramsay Shuck for their contributions to prior
iteration of the system.  We thank the reviewers for the helpful
feedback on the earlier revisions of the paper.  Finally, we thank Kelly Shaw
for her tireless shepherding and thoughtful suggestions on how to improve the
paper.

\bibliographystyle{IEEEtran}
\bibliography{locality}

\end{document}